\documentclass[10pt]{article}

\usepackage{latexsym,amssymb,eucal,graphicx}

\newtheorem{theorem}{Theorem}[section]
\newtheorem{proposition}{Proposition}[section]
\newtheorem{definition}{Definition}[section]

\newtheorem{lemme}{Lemma}[section]
\newtheorem{remark}{Remark}

\def\bbbr{\mathbb R}                     
\def\bbbc{\mathbb C}                     
\def\bbbn{\mathbb  N}                    
\def\bbbz{\mathbb  Z}                    
\def\P{\mathbb P}                        
\def\ii{\sqrt{-1}}                        
\def\A{\mathcal{A}}
\def\B{\mathcal{B}}
\def\C{\mathcal{C}}
\def\m{\mathfrak{m}}
\def\M{\mathcal{M}}                                    
\def\OO{ \mathcal{O}}                     
\def\L{\mathcal{L}}                       
\def\mequiv{\stackrel{\m}{\bf{\sim}}}
\def\T{{\bf T}}

\begin{document}

\roman{enumi}

\title{The monodromy of the  Lagrange top and the Picard-Lefschetz formula}
\author{Olivier Vivolo\\
\normalsize \it Laboratoire Emile Picard, U.R.A. C.N.R.S. 5580,\\
\normalsize \it  Universit\'e Paul Sabatier \\
\normalsize \it 118, route de Narbonne, 31062 Toulouse Cedex,
France.\\
\normalsize \tt vivolo@picard.ups-tlse.fr}

\maketitle
\begin{abstract}
The purpose of this paper is to show that the monodromy of action
variables of the Lagrange top and its generalizations can be deduced
from the monodromy of cycles on a suitable hyperelliptic curve (computed by the
Picard-Lefschetz formula).

\end{abstract}

\section{Introduction}

Let $(M,\omega)$ be a symplectic  manifold of dimension $2n$
and consider a Lagrangian fibration
$$
\begin{array}{cccl}
F: & M & \longrightarrow & B\\
\end{array}.
$$
where $B$ is a manifold of dimension $n$. We shall also suppose that 
each fiber
$F_q= F^{-1}(q)$ is compact and connected, so it is diffeomorphic to a
Liouville torus. 

For each $q\in B$ there is an open neighborhood $U\subset B$ of $q$ and a
diffeomorphism
 $$V= F^{-1}(U) 
\longrightarrow U \times \T^n : p \longmapsto
(I_1,...,I_n,\phi_1,...,\phi_n)$$ where $\T^n$ is the $n$-torus
$\bbbr^n/\bbbz^n$. Moreover the coordinates $I_i$, called action
coordinates, are smooth functions depending on $q$ only,
and in these coordinates the symplectic form is
$$\omega=\sum_{i=1}^{n}d\phi_i \wedge dI_i.$$
The coordinates
$\phi_i$ are called angle coordinates. 
Thus $V$ has the structure of a
symplectic principal bundle with a structure group $\T^n$,
Lagrangian fibers, and a Hamiltonian action of the structure group
whose momentum map is the projection map of the bundle. 

The question of global existence of action-angle coordinates on the
principal bundle
$M \longrightarrow  B$ 
has been studied in a pioneering paper by Duistermaat \cite{duistermaat}.
A most obvious obstruction to the global existence of such coordinates is of
course the monodromy of the bundle, which is a homomorphism from $\pi _1(B,b)$
to
$H_1(F_b,\bbbz) =
\bbbz^{n}$. The first example of a  mechanical system with non-trivial
monodromy
 is due to R. Cushman (the spherical pendulum, see \cite{duistermaat}).  It
turned out later that many other integrable systems have this property.
We mention here
 the Lagrange top \cite{cushman1}, the spherical
pendulum with quadratic potential $(x_3-a)^2$ \cite{zou2}, the
so called Kirchoff top (a rigid body in an infinite ideal fluid)
\cite{bates}.

General theorems in this direction are due to  
M. Zou \cite{zou} and T.Z. Nguyen
\cite{nguyen}. These results have an analytical nature: they do not use the
underlying algebro-geometric structure of the problem. In the present paper we 
shall develop this second (algebro-geometric) approach on a concrete example:
the Lagrange top and its generalizations. The idea of the proof is the
following. Let us suppose that we have an algebraically completely integrable
Hamiltonian system. This defines a Lagrangian fibration and we suppose that each
Lagrangian fiber (Liouville torus) can be complexified to an affine part of a
Jacobian variety
$J(\Gamma _b)= H^0(\Gamma _b,\Omega ^1)^*/H_1(\Gamma _b,\bbbz)$, where $\Gamma
_b$ is a spectral curve depending on $b$. The manifold $B$ is the complement
to the discriminant locus of the spectral curve
$\Gamma _b$. It is easier to describe the monodromy of the complexified
Lagrangian fibration (with fibers $J(\Gamma _b)$). Indeed, its monodromy
coincides with the monodromy of the homology Milnor bundle with fibers 
$H_1(\Gamma _b,\bbbz)$ and base $B$. We recall that the latter is associated
to the Milnor fibration of the polynomial defining the spectral curve
$\Gamma _b$. In particular it comes with a canonical Gauss-Manin connection
and its monodromy is computed by the Picard-Lefschetz theory (e.g.\cite{agv}).
Once the monodromy of the cycles of the homology Milnor bundle computed, it
remains to consider the monodromy of the  cycles generating the homology of
the real part of  $J(\Gamma _b)$, and hence of the real Liouville tori.

Of course if $B$ is simply connected there is no ({\em real }!) monodromy at
all. A simplified, but sufficiently general example is  when
$\Gamma _b$ is defined by a polynomial which itself is a versal deformation of
an isolated real simple singularity. The
complement to the real part of the complex discriminant locus may be not simply
connected (this set should not be confused with the complement to
the real discriminant locus, see \cite{looijenga}). The simplest non-trivial example is the
$A_3$ singularity $y^2 \pm x^4$ and the curve defined by its {\em real} versal
deformation is related to the spectral curve of the spherical pendulum
\cite{gavrilov3}. Indeed, the discriminant locus $\Delta_{ab}$ of the polynomial
$(x^2+1)^2+ax+b$ contains an isolated point ($a=0,b=0$).

The paper is organized as follows. In section \ref{s1} we define the
generalized Lagrange top as a $g+1$ degrees of freedom completely integrable
Hamiltonian system. The underlying algebro-geometric
structure is explained in section \ref{algebraic}. It turns out that, by analogy to
the classical Lagrange top ($g=1$) \cite{gavrilov1}, each  complexified
Liouville torus is an affine part of a generalized Jacobian 
$J(\Gamma')=H^0(\Gamma,\Omega ^1(\infty^+ +\infty^-)^*/H_1(\Gamma
_{aff},\bbbz)$ of a genus $g$
hyperelliptic curve $\Gamma $. Here $\Gamma_{aff}$ is a smooth
compact affine curve, $\Gamma $ is  the compactified and normalized $\Gamma_{aff}$,
$X \backslash \Gamma_{aff}= \infty^+ +\infty^-$, $\Gamma'$ is a compact singular curve obtained
from
$\Gamma $ by identifying $\infty^+$ and $\infty^-$. Therefore to compute the
monodromy of Liouville tori we have to determine first the monodromy of the homology bundle of
$\Gamma _{aff}$ (on the place of $\Gamma $), and then the monodromy of the cycles of 
$H_1(\Gamma
_{aff},\bbbz)$ which generate the homology of the real part of $J(\Gamma')$.
For this reason we need the real structure of $J(\Gamma')$ which is
described
in
section \ref{real}. Finally, using this and the Picard-Lefschetz formula,  we
compute the monodromy of the top, provided that
$g\leq 2$ (section \ref{last}).

This paper is an extended version of \cite{vivolo1}. I would like to thank 
Lubomir Gavrilov who suggested me the idea of the paper.

\section{Definition of the generalized Lagrange top}
\label{s1}
Consider the following Lax pair
\begin{equation}
\label{eq:system}
\frac{d}{dt} \Gamma(\lambda)=\left[\Gamma(\lambda),\chi \lambda+\Omega \right].
\end{equation}
where
$$\Gamma(\lambda)=\chi \lambda+ \Gamma_0
-\Gamma_1\lambda^{-1}\cdots-\Gamma_g\lambda^{-g}\in
\mathfrak{so}(3)[\lambda,\lambda^{-1}],  g \in \bbbn,$$ 
$$\chi=
\left(
  \begin{array}{ccc}
    0 & -1 & 0\\
    1 & 0 & 0\\
    0 & 0 & 0\\
  \end{array}
\right), \quad
\Omega=
\left(
  \begin{array}{ccc}
    0 & -\omega_3 & \omega_2\\
    \omega_3 & 0 & -\omega_1\\
    -\omega_2 & \omega_1 & 0\\
  \end{array}
\right),$$

$$
\Gamma_0=
\left(
  \begin{array}{ccc}
    0 & -(1+m)\omega_3 & \omega_2\\
    (1+m)\omega_3 & 0 & -\omega_1\\
    -\omega_2 & \omega_1 & 0\\
  \end{array}
\right), \quad i\in \{1,2,...,g\}\quad \Gamma_i= \left(
  \begin{array}{ccc}
    0 & -\gamma_{i,3} & \gamma_{i,2}\\
    \gamma_{i,3} & 0 & -\gamma_{i,1}\\
    -\gamma_{i,2} & \gamma_{i,1} & 0\\
  \end{array}
\right). $$

To simplify the notations we note below
$$\gamma_{0,1}=\omega_1, \quad \gamma_{0,2}=\omega_2, \quad
\gamma_{0,3}=(1+m)\omega_3.$$ 
The Lax pair  (\ref{eq:system}) has $2g+2$ first integrals
$$
H_k=-\frac{1}{4}\mbox{residue}_{\lambda=0}(\lambda^{k-1} tr
(\Gamma(\lambda)^2)), k= -1,0,1,...,2g.
$$ 
We have in particular 
$$H_{-1}=(1+m)\omega_3, \quad
H_0=\frac{1}{2}\left( \omega_1^2+\omega_2^2+(1+m)^2\omega_3^2
\right) -\gamma_{1,3}.$$ 
The Lax pair (\ref{eq:system}) can be written in an equivalent form as a
Hamiltonian system
$$
\frac{d}{dt}x=\{ x, H\}, 
$$
where
$$
H=H_0-\frac{m}{2(1+m)}H_{-1}^2=\frac{1}{2}\left(
\omega_1^2+\omega_2^2+(1+m) \omega_3^2 \right)-\gamma_{1,3}.
$$
The Poisson
structure $\{.,.\}$ is
given by
\begin{equation}
\label{eq:structure Poisson} \{
\gamma_{i,k},\gamma_{j,l}\}=\sum_{c=1}^3
\Lambda_{kl}^{c}\gamma_{{i+j},c} \quad \mbox{for }i,j\in \{
0,1,...,g\}.
\end{equation}
where $\Lambda_{kl}^{c}$ is a skew-symmetric matrix
$$\Lambda_{12}=\left(
\Lambda_{12}^{1},\Lambda_{12}^{2},\Lambda_{12}^{3}
\right)=(0,0,-1), \quad \Lambda_{13}=\left(
\Lambda_{13}^{1},\Lambda_{13}^{2},\Lambda_{13}^{3}
\right)=(0,1,0), \quad \Lambda_{23}=\left(
\Lambda_{23}^{1},\Lambda_{23}^{2},\Lambda_{23}^{3}
\right)=(-1,0,0).$$

It is easy to check further that
(\ref{eq:system}) is a Liouville completely integrable Hamiltonian system of
$g+1$ degrees of freedom, where $H_i$, $i=-1,0,...,g-1$ are first integrals,
while $H_j$, $j=g,g+1,...,H_{2g}$ are Casimirs.

We call the system (\ref{eq:system}) {\bf the generalized Lagrange top} (another
generalization may be found in \cite{reyman}).

We shall identify the Lie algebras $(\mathfrak{so}(3),[.,.])$ and
$(\bbbr^3,\wedge)$ by the Lie algebras anti-isomorphism
($[A,B]=-A\wedge B)$
$$ \left(
  \begin{array}{ccc}
    0 & -x_3 & x_2\\
    x_3 & 0 & -x_1\\
    -x_2 & x_1 & 0\\
  \end{array}
\right) \in \mathfrak{so}(3) \longmapsto (x_3,x_2,x_1)\in
\bbbr^3.$$

Let $ \sigma_1,\sigma_2,\sigma_3 $ be the Pauli spin matrices, defined by

\[
\sigma_1=
  \left(
  \begin{array}{cc}
    1 & 0 \\
    0 & -1
  \end{array}
  \right), \quad
\sigma_2 =\ii
  \left(
  \begin{array}{cc}
    0 & -1 \\
    1 & 0
  \end{array}
  \right), \quad
\sigma_3=
  \left(
  \begin{array}{cc}
    0 & 1 \\
    1 & 0
  \end{array}
  \right),
\]
and denote $\sigma=(\sigma_1,\sigma_2,\sigma_3).$ Then
$[\sigma_1,\sigma_2]=2\ii \sigma_3$ (+cyclic permutation)
which implies that the map $$x=\left( x_{3}, x_{2}, x_{1}
\right) \in \bbbr^3 \longmapsto
\frac{1}{2\ii}\widetilde{x}=\frac{1}{2\ii} x.\sigma=\frac{1}{2}
\left(
\begin{array}{cc}
    -\ii x_1 & -\ii x_3 -x_2 \\
    -\ii x_3 +x_2 & \ii x_1
  \end{array}
  \right)\in \mathfrak{su}(2), $$
where $\widetilde{x}=x.\sigma=x_1\sigma_1+x_2\sigma_2+x_3\sigma_3$
is a Lie algebra isomorphism between $\bbbr^3$ and the $(2\times
2)$ skew-Hermitian traceless matrices $\mathfrak{su}(2)$. Note
that 
$$-\det (x.\sigma)=||x||^2, \quad \mbox{and
trace}(\widetilde{x}\widetilde{y})=-\frac{1}{2}x.y.$$

Composing these two previous morphisms of Lie algebras we get a
Lie algebras anti-isomorphism between $(\mathfrak{so}(3),[.,.])$
and $(\mathfrak{su}(2),[.,.])$, we deduce from (\ref{eq:system}) an
equivalent Lax  pair. Namely,
$$ \mathfrak{so}(3) \ni \chi 
\longmapsto
\frac{1}{2\ii}\sigma_3 \in \mathfrak{su}(2),$$ $$
\mathfrak{so}(3) \ni \Omega 
\longmapsto \frac{1}{2\ii} \widetilde{\Omega}=
\frac{1}{2\ii}\left(\begin{array}{cc} \omega_1 & \omega_3-\ii
\omega_2\\
 \omega_3+\ii \omega_2 & -\omega_1
\end{array}\right)\in \mathfrak{su}(2),$$ and finally 
$$
 \mathfrak{so}(3) \ni \Gamma_i  \longmapsto
\frac{1}{2\ii}\widetilde{\Gamma}_i=\frac{1}{2\ii}\left(\begin{array}{cc}
\gamma_{i,1} & \gamma_{i,3}-\ii \gamma_{i,2} \\ \gamma_{i,3}+\ii
\gamma_{i,2} & -\gamma_{i,1}
\end{array}\right)\in \mathfrak{su}(2), i=1,2,...,g.$$
If we denote
 $$U(x)=x^{g+1}+\left((1+m)\omega_3-\ii \omega_2\right)x^g-
\left(\gamma_{1,3}-\ii \gamma_{1,2}
\right)x^{g-1}-\cdots-(\gamma_{g,3}-\ii \gamma_{g,2}), $$
$$W(x)=x^{g+1}+\left((1+m)\omega_3+\ii \omega_2\right)x^2-
\left(\gamma_{1,3}+\ii \gamma_{1,2}
\right)x^{g-1}-\cdots-(\gamma_{g,3}+\ii \gamma_{g,2}),$$
$$V(x)=\omega_1 x^g-\gamma_{1,1}x^{g-1}-\gamma_{2,1}
x^{g-2}-\cdots-\gamma_{g,1}, $$ then
$$\Gamma(\lambda) \longmapsto
\frac{1}{2\ii}\widetilde{\Gamma}(x)=\frac{1}{2\ii}
\left(\begin{array}{cc} V(x) & U(x) \\ W(x) & -V(x)
\end{array}\right)=\frac{1}{2\ii}\left(
\sigma_3 x^{g+1}+
\widetilde{\Gamma}_{0}x^g-\widetilde{\Gamma}_{1}x^{g-1}-\cdots-\widetilde{\Gamma}_g
\right).$$ The generalized Lagrange top (\ref{eq:system}) becomes
under this anti-isomorphism
$$2\ii\frac{d}{dt}\widetilde{\Gamma}(x)=\left[\sigma_3x+\widetilde{\Omega},\widetilde{\Gamma}(x)\right].$$
In the next section we shall describe the algebro-geometric structure of the
complexified generalized Lagrange top. Therefore we put 
$(x_3,x_2,x_1)\in \bbbc^3$ and consider the Lie
algebra anti-isomorphism between ($\mathfrak{so}(3,\bbbc),[.,.]$)
and ($\mathfrak{sl}(2,\bbbc), [.,.]$).

\section{Algebraic structure}
\label{algebraic}

In this section, we show that the generalized Lagrange top is an
algebraically completely integrable system in the sense of
Mumford \cite[p.3.53]{mumford}. This means that the generic complex level set
of this system is an affine part of a commutative algebraic group : the
generalized Jacobian $J(C,\infty^{\pm})$ of an hyperelliptic curve
of genus $g$ with two points $\infty^{\pm}$ identified.

The construction and properties of generalized Jacobians are due
to Rosenlicht \cite{rosenlicht1,rosenlicht2} (even if the
generalized Jacobian have been already used by Jacobi
\cite{jacobi}) and Lang \cite{lang1,lang2}; they rely on the
theory of abelian varieties, developed by Weil \cite{weil}.

Below we shall use the Serre's notations \cite{serre}.

Let $C$ be the compact and normalized hyperelliptic curve defined by
equation
$y^2=f(x)=\prod_{i=1}^{2g+2}(x-x_i)$. Let $\iota$ be the
hyperelliptic involution $\iota : (x,y)\in C \longmapsto (x,-y)\in
C$. Denote by $\infty^+$, $\infty^-$, the two points "at infinity" on $C$
 ($\infty^+=\iota (\infty^- $)),
and $\breve{C}=C\backslash \{\infty^+,\infty^-\}$. 
The pair $(C,\infty^\pm)$ defines a singular curve $C'$ (the singularization
of $C$ with respect to the modulus $\infty^++\infty^-$). As a topological
space $C'$ is $C$ with the two points $\infty^+,\infty^-$ identified.
 The structure sheaf $\OO'$ of $C'$ is defined
in the following way. Let $\OO_{C'}$ be the direct image of the
structure sheaf $\OO_C$ under canonical projection $C
\longrightarrow C'$. Then $$\OO_P^{'}=\left\{
\begin{array}{ll}
\OO_P & \mbox{if }P \in \breve{C}\\
\bbbc+i_\infty & \mbox{if }P =\infty\\
\end{array}
\right.$$
where $i_\infty$ is the ideal of $\OO_\infty$ formed by the functions $f$ having a
zero at $\infty^+$ and $\infty^-$ of order at least $1$.
We define the sheaf $\L^{'} (D)$ where $D$ is a divisor on $C$ such that $\mbox{Supp}(D)\bigcap
\{\infty^+,\infty^- \}=\emptyset$
by
$$\L^{'}(D)_P=\left\{
\begin{array}{ll}
\ \L(D)_P & \mbox{if }P \in \breve{C}\\
\ \OO^{'}_\infty & \mbox{if }P =\infty\\
\end{array}
\right.$$
Let
$$\left.
\begin{array}{ll}
\ L'(D)=H^0(C',\L^{'}(D)), & I^{'}(D)=H^1(C^{'},\L^{'}(D)),\\
\ l'(D)=\dim_\bbbc L^{'}(D), & i'(D)=dim_\bbbc I^{'}(D).\\
\end{array}
\right.$$ As the sheaf $\OO_C/\OO_C^{'}$ is coherent, let
$\delta_P=\mbox{dim}_\bbbc(\OO_P/\OO_P^{'})$ with $P \in C'$, the
arithmetic genus $p_a$ (dimension of $H^1(C',\OO')$) of the
singular curve $C'$
 is obtained from the geometric genus $g$ of $C$ by the relation
$$p_a=g+\delta_\infty.$$
In fact
$$\delta_\infty=\dim_\bbbc(\OO_\infty/(\bbbc+i_\infty))=
\dim_\bbbc\left(\OO_\infty/i_\infty\right)-1=\mbox{deg}(\m)-1=1$$
then $$p_a=g+1.$$
A divisor $D$ on $C$ with $\mbox{Supp}(D)\bigcap
\{\infty^+,\infty^- \}=\emptyset$ verifies
$$l'(D)-i'(D)=\mbox{deg}(D)+1-p_a=\mbox{deg}(D)-g.$$
Now we define the equivalence relation $\mequiv$.

\begin{definition}
Let $D_1$ and $D_2$ be two divisors on $C$ with
$\mbox{Supp}(D_1)\bigcap \{\infty^+,\infty^- \}=\emptyset$ and
$\mbox{Supp}(D_2)\bigcap \{\infty^+,$ $ \infty^- \}=\emptyset$. Then $D_1
\mequiv D_2$
provided that there exists a global meromorphic function $f$ on
$C$, such that $(f)=D_1-D_2$ and $v_{\infty^{\pm}}(f-1)\geq 1$.
\end{definition}

\begin{definition}
The generalized Jacobian of $C'$, denoted $J(C,\infty^\pm)$, is the subgroup
$\mbox{Pic}^0(C')$ of
$\mbox{Pic}(C'):=\mbox{Div}(C')/\mequiv$
formed by the divisors $D$ on $C$ with $\mbox{Supp}(D)\bigcap \{\infty^+,\infty^- \}=\emptyset$
and $\mbox{deg}(D)=0$.
\end{definition}

It is known that $J(C,\infty^\pm)$ is an extension of $J(C)$ the usual Jacobian of $C$ by the algebraic group
$\bbbc^*$:
$$0 \longrightarrow {\bbbc^*}\longrightarrow J(C,\infty^\pm)
\stackrel{\phi}{\longrightarrow}
J(C) \longrightarrow 0$$

An explicit embedding of a Zariski open subset of
$J(C,\infty^{\pm})$ in $\bbbc^{3(g+1)}$ is constructed by the
following classical construction due to Jacobi \cite{jacobi} and
Mumford \cite{mumford}. Let
$$f(x)=x^{2g+2}+a_{1}x^{2g+1}+a_{2}x^{2g}+\cdots+a_{2g+2}$$ be a
polynomial without double roots and define the Jacobi polynomials
$$U(x)=x^{g+1}+u_g x^g+u_{g-1} x^{g-1}+\cdots+u_0, \quad
V(x)=v_{g}x^g+v_{g-1}x^{g-1}+\cdots+v_0,$$ $$W(x)=x^{g+1}+w_g
x^g+w_{g-1} x^{g-1}+\cdots+w_0.$$ Let $T_C$ be the set of Jacobi
polynomials satisfying the relation $$f(x)=V^2(x)+U(x)W(x).$$

More explicitly, if we expand
 $$f(x)-V^2(x)-U(x)W(x)=\sum_{i=0}^{2g+1}
c_i(a_j,u_k,v_l,w_m) x^{i},$$ and take
$u_j,v_k,w_l$ as coordinates in $\bbbc^{3(g+1)}$, then
$$T_C=\left\{(u,v,w)\in \bbbc^{3(g+1)}: \quad
c_i(a_j,u_k,v_l,w_m)=0, \quad i\in \{0,1,...,2g+1\} \right\}.$$

\begin{proposition}
\label{propos} If $f(x)$ is a polynomial without double root then
\begin{enumerate}
  \item $T_C$ is a smooth affine variety isomorphic to $J(C,\infty^{\pm})\setminus \Theta$ for some divisor
  theta. Under $\phi$, the set $\Theta$ is the translate of the set
  of special divisors of degree $g-1$ by $\infty^+ +\infty^-$.
  \item any translation invariant vector field on the generalized Jacobian of the
  curve $C$ with modulus $\m=\{\infty^+,\infty^-\}$, can be written in the following Lax pair form
$$2\ii
\frac{d}{dt}\widetilde{\Gamma}(x)=\left[\widetilde{\Gamma}(x),\frac{\widetilde{\Gamma}(a)}{x-a}
\right],\quad \quad \widetilde{\Gamma}(x)=\left(\begin{array}{cc}
V(x) & U(x)
\\ W(x) & -V(x) \end{array}\right), $$ where $a\in \bbbc$ and
$U(x),V(x),W(x)$ are the Jacobi polynomials.

\end{enumerate}

\end{proposition}

\paragraph{Proof} The proof of part (1) of the above proposition
can be found in Previato \cite{previato}. For the proof of part
(2) see
\cite{beauville,gavrilov2,gavrilov1}.\\

Let $Div^{g+1}(\breve{C})$ be the set of positive divisors of
degree $g+1$ on $\breve{C}$ and $Div^{+,g+1}_0(\breve{C})\subset
Div^{g+1}(\breve{C})$ be the subset of divisors $\displaystyle
D=\sum_{i=1}^{g+1} P_i$ on $\breve{C}$ having the property
$Supp(D) \cap Supp(\iota(D))=\emptyset$. The set
$Div^{+,g+1}_0(\breve{C})$ is naturally identified with a Zariski
open subset of the symmetric product $S^{g+1}\breve{C}$. There is
a bijection between $T_C$ and $Div^{+,g+1}_0(\breve{C})$. In fact
$T_C$ is smooth and the bijection is an isomorphism of smooth
algebraic varieties \cite{mumford}.

For some fixed divisor $\displaystyle D_0=\sum_{i=1}^{g+1} W_i\in
Div^{+,g+1}_0(\breve{C})$, we consider the Abel-Jacobi map $$
 \begin{array}{crcl}
\displaystyle
  \A : & Div^{+,g+1}_0(\breve{C})\subset S^{g+1}\breve{C} & \longrightarrow & J(C,\infty^{\pm}) \\
   & D=\sum_{i=1}^{g+1} P_i & \longmapsto &  \int_{D_0}^D \omega:=\left( \sum_{i=1}^{g+1}\int_{W_i}^{P_i}
dx/y ,\sum_{i=1}^{g+1}\int_{W_i}^{P_i} xdx/y,...,\sum_{i=1}^{g+1}
\int_{W_i}^{P_i} x^g dx/y \right).
\end{array}
$$ Next we apply the proposition \ref{propos} to generalized
Lagrange top. Let $C_{\underline{h}}$ be the curve $C$ as above,
where $\underline{h}=(h_{-1},h,h_1,...,h_{2g})\in \bbbc^{2(g+1)}$,
and $$
f(x)=x^{2g+2}+2h_{-1}x^{2g+1}+2hx^{2g}+2h_1x^{2g-1}+\cdots+2h_{2g}.$$
Let us consider the complex invariant level set of the generalized
Lagrange top

$$T_{\underline{h}}=\left\{(\omega_i,\gamma_{j,k})\in
\bbbc^{3(g+1)}: H_{-1}(\omega_i,\gamma_{j,k})=h_{-1},H
(\omega_i,\gamma_{j,k})=h, H_1(\omega_i,\gamma_{j,k})=h_1,...,
H_{2g}(\omega_i,\gamma_{j,k})=h_{2g}\right\}.$$
This linear change
of variables
\begin{equation}
\label{eq:linearchange}
\left\{
\begin{array}{rcl}
u_g & = & (1+m)\omega_3-\ii\omega_2\\ u_{g-1} & =
&-\gamma_{1,3}+\ii \gamma_{1,2}\\ u_{g-2} & = & -\gamma_{2,3}+\ii
\gamma_{2,2}\\ ...\\ u_{0} & = & -\gamma_{g,3}+\ii \gamma_{g,2}\\
\end{array}
\right.
, \quad
\left\{
\begin{array}{rcl}
v_g & = & \omega_1\\ v_{g-1} & = & -\gamma_{1,1}\\ v_{g-2} & = &
-\gamma_{2,1}\\ ...\\ v_0 & = & -\gamma_{g,1}\\
\end{array}
\right.
, \quad
\left\{
\begin{array}{rcl}
w_g & = & (1+m)\omega_3+\ii\omega_2\\ w_{g-1} & = &
-\gamma_{1,3}-\ii \gamma_{1,2}\\ w_{g-2} & = & -\gamma_{2,3}-\ii
\gamma_{2,2}\\ ...\\ w_0 & = & -\gamma_{g,3}-\ii \gamma_{g,2}\\
\end{array}
\right. ,
\end{equation}

identifies $T_C$ and $T_{\underline{h}}$ where the curves $C$ and $C_{\underline{h}}$ are related in the
following way
$$a_{1}=2h_{-1}, \quad a_{2}=2h=2h_0-\frac{m}{1+m}h_{-1}^2, \quad a_{3}=2h_1, \quad ..., \quad a_{2g+1}= 2h_{2g-1}, \quad
    a_{2g+2}=2h_{2g}.$$

We summarize this in the following
\begin{theorem}

\begin{enumerate}
  \item The complex level set $T_{\underline{h}}$ is a smooth complex manifold
bi holomorphic to
  $J(C,\infty^{\pm})\setminus \Theta$ where $\Theta$ is a theta divisor $\left(\Theta=J(C,\infty^{\pm})
  \setminus \A\left(Div^{+,g+1}_0(\breve{C})\right)\right)$.

  \item The Hamiltonian flows of generalized Lagrange top restricted to $T_{\underline{h}}$ induce linear flows
  on $J(C,\infty^{\pm})$. The corresponding vector fields $\{\;.\; ,H_{-1} \},\{\;.\; ,H \}, \{\;.\; ,H_i \}$
  for $i\in \{1,2,...,g-1\}$ have a Lax pair
  representation obtained from the Lax pair (\ref{eq:lax pair}) by substituting $a\in
  \P^1$ and using the linear change of variables
  (\ref{eq:linearchange}).

\begin{equation}
\label{eq:lax pair} 2\ii
\frac{d}{dt}\widetilde{\Gamma}(x)=\left[\widetilde{\Gamma}(x),\frac{\widetilde{\Gamma}(a)}{x-a}
\right].
\end{equation}

\end{enumerate}

\end{theorem}

\section{The Real Structure}
\label{real}
Consider the set $\bbbr^{2(g+1)}$ of all real polynomials of the form
$f(x)=x^{2g+2}+2h_{-1}x^{2g+1}+2hx^{2g}+2h_1x^{2g-1}+\cdots+2h_{2g}$.
its coefficients are real and its roots are distinct. Denote by 
$\Delta \subset \bbbr^{2(g+1)}$ its discriminant locus. Denote further by
$\C $ the connected component of the complement to $\Delta $ in 
$\bbbr^{2(g+1)}$, in which $f(x)$ has no real root (obviously there is only one
such component).

We recall that a real structure on a complex algebraic variety $C$
is an anti-holomorphic involution $S : C \longrightarrow C$
(e.g.\cite{silhol}). The real structure on
$T_{\underline{h}}$ is given by the usual complex conjugation
$$(\omega_i,\gamma_{1,j},\gamma_{2,k},...,\gamma_{g,l}) \longmapsto
(\overline{\omega_i},\overline{\gamma_{1,j}},\overline{\gamma_{2,k}},...,\overline{\gamma_{g,l}})$$
and we denote $\displaystyle
T_{\underline{h}}^{\bbbr}:=T_{\underline{h}} \cap \bbbr^{3(g+1)}$.

There are two natural  anti-holomorphic involution on
$J(C,\infty^\pm)\backslash \Theta$ $$J_1: (U,V,W)\longrightarrow
(\overline{U},-\overline{V},\overline{W}), \quad J_2:
(U,V,W)\longrightarrow (\overline{W},\overline{V},\overline{U}).$$
Denote by $\M_1$, (respectively $\M_2$ ) the set of fixed points
of $J_1$ ($J_2$)
$$\M_1=\left\{(U,V,W): \quad U,V \mbox{ real}, V \mbox{ imaginary}
\right\},$$ 
$$\M_2=\left\{(U,V,W): \quad U=\overline{W}, V \mbox{
real} \right\}.$$

\begin{proposition}
The real structure on $T_C$ is given by the involution $J_2$ and
$\M_2=T_{\underline{h}}^{\bbbr}$.
\end{proposition}
\paragraph{Proof} 
Fixed points of $J_2$ in $T_{C}$ give real
$(\omega_i,\gamma_{1,j},\gamma_{2,k},...,\gamma_{g,l}) $ and
vice versa. \\ \\

\begin{figure}
\centering
\includegraphics[width=9cm,height=6cm]{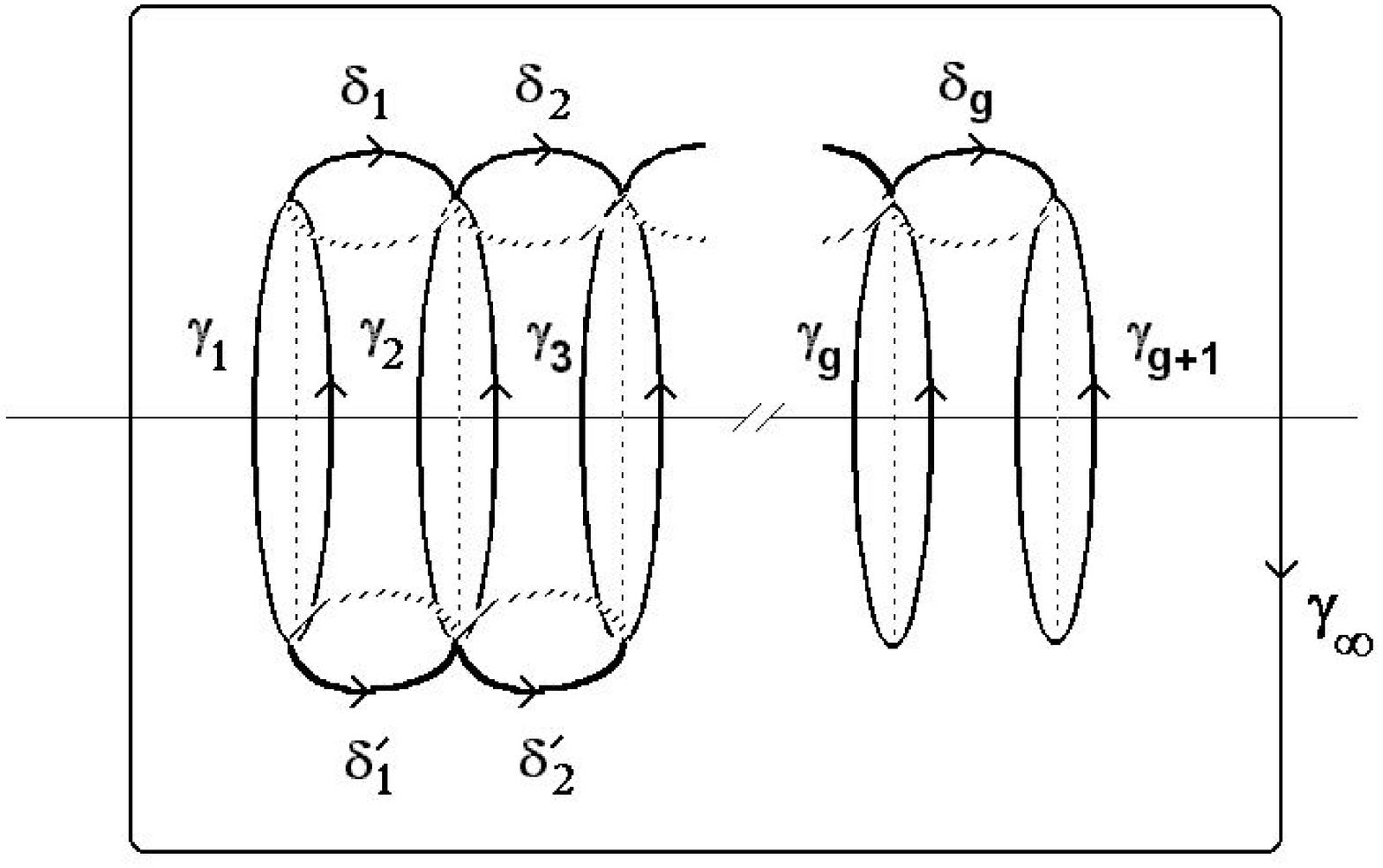}
\caption{Projection of the cycles $\delta_i$ and $\gamma_j$ on the
$x$-plane} \label{fig:cycles}
\end{figure}

Let $W_i$ be $2(g+1)$ Weierstrass points on $\breve{C}_h$, where (without loss
of generality) we suppose that
$\sum_{i=1}^{g+1}W_i=\sum_{i=1}^{g+1}\overline{W_{g+1+i}}$. Let us
choose a basis $\{\gamma_i,\delta_j\}_{i\in
\{1,...,g+1\},j\in\{1,...,g\}}$ of $H_1(\breve{C},\bbbz)$ as shown
on figure \ref{fig:cycles}. Given $\omega=\left( dx/y, x
dx/y,...,x^g dx/y\right)$, and $e_i=\oint_{\gamma_i} \omega,
i=1,2,...,g+1$, $f_j=\oint_{\delta_j}  \omega, j=1,2,...,g$, we
define  $\Lambda_{2g+1}$ to be the $\bbbz$-module  $\bbbz
\{e_1,...,e_{g+1},f_1,...,f_g \}$.

\begin{proposition}
Assume that $f(x)$ is a real polynomial with simple roots.
\begin{enumerate}
\item $T_{\underline{h}}^{\bbbr}$ is not empty if and only if $\underline{h}\in \C$.
\item
The real structure $J_2$ acts on $J(C,\infty^{\pm})$ as $z \in
\bbbc^{g+1} / \Lambda_{2g+1} \longmapsto -\overline{z} \in
\bbbc^{g+1} / \Lambda_{2g+1} $, where $\overline{z}$ is the
complex conjugation on $\bbbc^{g+1}$.
\end{enumerate}
\end{proposition}
{\bf Proof} The definition of $J_2$ gives that if $(U,V,W)\in T_C$
and $J_2(U,V,W)=(U,V,W)$ then
$$V^2(x)+U(x)W(x)=|V(x)|^2+|U(x)|^2=f(x)\geq 0, \quad \forall x\in
\bbbr.$$ In
If $f(x)$ vanishes then this zero is in fact double, and this is
impossible.  This shows that $f(x)$ is strictly positive.
 Reciprocally, if
$\underline{h}\in \C$ then $J_2(\sum_{i=1}^{g+1}
W_i)=\sum_{i=1}^{g+1} W_{i}$.
\\
\\
Now let us determine the action of $J_2$ on $S^{g+1} \breve{C}$.
Let $P_1=(x_1,y_1),P_2=(x_2,y_2),...,P_{g+1}=(x_{g+1},y_{g+1})$ be
generic points on $\breve{C}$. Let us consider the curve
$X=\{(x,y)\in \bbbc^2 : \quad y=V(x) \}$ where $V(x)$ is the
Lagrange polynomial of degree $g$ such that $X$ contains
$P_1,P_2,...,P_{g+1}$. The intersection points between $\breve{C}$
and $X$ are the points $P_1,P_2,...,P_{g+1},Q_1,...,Q_{g+1}$. The
points $Q_i=(x^i,y^i)$ are determined simply by $y^i=V(x^i)$ where
$x^i$ are roots of polynomial $V^2(x)-f(x)$ (which is the
resultant of $y-V(x)$ and $y^2-f(x)$ with respect to $y$). We have
\begin{center}
$((y-V(x))|_C)=D_1+D_2-(g+1)(\infty^++\infty^-)$ where
$\displaystyle D_1=\sum_{i=1}^{g+1}P_i$ and $\displaystyle
D_2=\sum_{i=1}^{g+1} Q_i$,
\end{center}
\begin{center}
$(y)=D_0+D'_0-(g+1)(\infty^++\infty^-)$ where $\displaystyle
D_0=\sum_{i=1}^{g+1}W_i$ and $\displaystyle
D'_0=\sum_{i=1}^{g+1}W_{g+1+i}$.
\end{center}
We get $((y-V(x))/y |_C)=D_1+D_2-D_0-D'_0$ and $(y-V(x))/y
(\infty^{\pm})=1$ then $\displaystyle D_1-D_0 \mequiv D'_0-D_2$.
Choose $D_0=\sum_{i=1}^{g+1}W_i$ as the base point of the Abel-Jacobi
map $\A$.

Recall that if 
 $S$ is real structure on $C$, then $S$ induces a transformation
on the sheaves ${\mathcal O}_C$, $\Omega^1$, $\bbbz$. $$
\begin{array}{crcl}
   S^*: & \Gamma(U, {\mathcal O}_C) & \longrightarrow &  \Gamma(S(U),{\mathcal O}_C)\\
     & f & \longmapsto &  \overline{f o S}\\
  \end{array}$$
 We also denote by $S^*$ the transformation induced on $\Omega^1$. We
shall say that $\alpha \in  H^0(C,\Omega^1)$ is $S$-real provided that $S^*
\alpha =\alpha $. Moreover $S$ induces an involution on $C_1(C,\bbbz)$ (the
group of topological 1-cycles). If $\alpha  \in H^0(C,\Omega^1)$ and $c\in
C_1(C,\bbbz)$ then $\displaystyle \int_c S^*
\alpha =\overline{\int_{S(c)}\alpha }$. If $\alpha $ is $S$-real, we
get $\displaystyle \int_c \alpha =\overline{\int_{S(c)}\alpha }$. We
shall say $c \in H_1(C,\bbbz)$ is $S$-real ($S$-imaginary)
if $S(c)=c$ ($S(c)=-c$ ).

The differential one-forms $x^idx/y$ on $C$ are real (for
the usual real structure), and if we denote
$\omega=\left(dx/y,xdx/y,...,x^gdx/y\right)$ then
 $$\int_{D_0}^{J_2(D_1)}
\omega=\overline{\int_{D'_0}^{D_2} \omega}=-\overline{
\int_{D_2}^{D'_0} \omega} =-\overline{\int_{D_0}^{D_1} \omega}.$$
Therefore the involution $J_2$ acts on $J(C,\infty^{\pm})$ as $z
\longmapsto J_2(z)=-\overline{z}$, $z \in \bbbc^{g+1}/
\Lambda_{2g+1}$, where $\displaystyle z=\int_{D_0}^{D_1} \omega $
and $\displaystyle J_2(z)=\int_{D_0}^{J_2(D_1)} \omega$.
\begin{theorem}
$T_{\underline{h}}^{\bbbr} \subset \bbbc^{g+1} /\Lambda_{2g+1}$ is
topologically a $(g+1)$-torus and its periods are generated by
$e_i, \quad i\in \{1,2,...,g+1\}$.
\end{theorem}
{\bf Proof } The fact that
$T_{\underline{h}}^{\bbbr}$ is compact and connected is proved by Previato
\cite{previato}. Consider the image of
$T_{\underline{h}}^{\bbbr}$ in $J(C,\infty^{\pm})$ under the
Abel-Jacobi map. As $\omega$ is real and $\gamma_i$ are
imaginary cycles, then $e_i\in \bbbc^{g+1}$ are purely imaginary
 vectors. We shall determine the action of $J_2$ on
$H_1(\breve{C}_h,\bbbz)$ and hence on the period lattice
$\Lambda_{2g+1}$. Let us choose a base of $H_1(\breve{C}_h,\bbbz)$
as on figure \ref{fig:cycles}.

Under the standard anti-holomorphic involution $\delta_j$ is sent to
$\delta'_j$ which is homologous to $\displaystyle
\delta_j-\gamma_{\infty}-\sum_{i=1,\\i\neq j,\\i\neq j+1}^{g+1}
\gamma_i$. As $\displaystyle \gamma_{\infty}\equiv
-\sum_{i=1}^{g+1} \gamma_i$ then $\delta_j'\equiv
\delta_j+\gamma_j+\gamma_{j+1}$. Thus
\begin{center}
$\overline{f_j}=f_j+e_j+e_{j+1}$ and
$J_2(f_j)=-\overline{f_j}=-f_j-e_j-e_{j+1} $. \end{center} Denote
by
$z\in \bbbc^{g+1}$, $\Re(z) \in \bbbr^{g+1}$ the real part of
$z$.\\ \\ Complete further $\{e_1,...,e_{g+1},f_1,...,f_g\}$ to a basis of
$\bbbc^{g+1}$  by $\{e_1,...,e_{g+1},f_1,...,f_g,f_{g+1}\}$
under the condition  that\\
$\{ \Re{(f_1)},... , \Re{(f_{g+1})}, f_{g+1} \}$
is a basis of $\bbbr^{g+1}$. The fixed points
of $J_2$ in $\bbbc^{g+1}$ are given by $$J_2z=z, \mbox{ where }
J_2= \left (
\begin{array}{cc}
Id_{g+1} & -A\\ 0 &-Id_{g+1}\\
\end{array}
\right) \mbox{ and }
A=\left (
\begin{array}{cccccc}
1 & 0 & 0 & \ldots &0 &0\\
1 & 1 & 0 & \ldots &0 &0\\
0 & 1 & 1 & \ldots &0 &0\\
\vdots  &   & 1 & \ldots &0 &0\\
  &   &   &    &1 &0\\
0  & \ldots  &   &    &1 &0\\
\end{array}
\right).$$ If  $z\in \bbbc^{g+1}/{\bbbz
\{e_1,...,e_{g+1},f_1,...,f_{g+1}\}}$ the only possible solutions are
\begin{center}
$\forall (q_1,...,q_{g+1}) \in S^{g+1}, \quad \forall j\in \{
1,2,...,g\} \quad p_j \equiv 0 \mbox{ mod } f_j \mbox{ and } \quad
2p_{g+1} \equiv 0 \mbox{ mod } f_{g+1}$.
\end{center}
Assume that the vector $f_{g+1}$ tends  to infinity,
 and get
\begin{center}
$\forall (q_1,...,q_{g+1}) \in S^{g+1},\quad  \forall j\in \{
1,2,...,g\}, \quad p_j \equiv 0 \mbox{ mod } f_j$.
\end{center}
Finally $T_{\underline{h}}^{\bbbr}$ is generated by $e_i$ for
$i\in \{1,2,...,g+1\}$.

\section{The Monodromy}
\label{last}
\subsection{The case $g=0$}
 The system (\ref{eq:system}) is
$$\frac{d}{dt}\Gamma_0=\left[\Gamma_0,\Omega \right]$$ or equivalently
$$
\displaystyle \left\{
  \begin{array}{ccc}
    \dot{\omega_1} & = & - m \omega_2 \omega_3\\
    \dot{\omega_2} & = & m \omega_1 \omega_3\\
    \dot{\omega_3} & = & 0\\
  \end{array}
\right.
$$
It is a Hamiltonian system with one degree of freedom with 
Poisson structure
$$
  \begin{array}{c||ccc}
    \{.,.\} & \omega_1 & \omega_2 & \omega_3 \\
    \hline \hline
    \omega_1 & 0 & -(1+m)\omega_3 & \omega_2/(1+m) \\
    \omega_2 & (1+m)\omega_3 & 0 & -\omega_1/(1+m) \\
    \omega_3 & -\omega_2/(1+m) & \omega_1/(1+m) & 0 \\
  \end{array}
$$
and 
Hamiltonian 
$$H=H_0-\frac{m}{2(1+m)}H_{-1}^2=\frac{1}{2}\left(
\omega_1^2+\omega_2^2+(1+m) \omega_3^2 \right),$$ where
$$H_{-1}=(1+m)\omega_3$$ 
is a first integral and
$$H_{0}=\frac{1}{2}\left( \omega_1^2+\omega_2^2+(1+m)^2\omega_3^2
\right)$$
is a Casimir function.
 The spectral curve associated to the Lax pair
(\ref{eq:lax pair}) is given by the polynomial
$$y^2-f(x)=y^2-U(x)W(x)-V^2(x)=y^2-x^2-2h_{-1}x-\left(2h+
\frac{m}{1+m}h_{-1}^2\right)=0.$$
It is a genus zero curve and its generalized Jacobian is $\bbbc^*$. It is 
identified to the invariant manifold of the system. The spectral curve as
well the corresponding Lagrangian fibration have no monodromy.

\subsection{The case $g=1$ (the Lagrange top)}

The system (\ref{eq:system}) is
\begin{equation}
\label{eq:lagrange top}
\frac{d}{dt}\left( \chi \lambda +\Gamma_0-\Gamma_1\lambda^{-1} \right)=\left[\chi \lambda
+\Gamma_0-\Gamma_1\lambda^{-1},\chi
\lambda +\Omega \right].
\end{equation}
or equivalently
$$\frac{d}{dt}\Gamma_0=\left[\Gamma_0,\Omega
\right]-\left[\Gamma_1,\chi \right],$$
$$\frac{d}{dt}\Gamma_1=\left[\Gamma_1,\Omega \right].$$ 
If we denote
$$\Gamma_1=
\left(
  \begin{array}{ccc}
    0 & -\gamma_3 & \gamma_2\\
    \gamma_3 & 0 & -\gamma_1\\
    -\gamma_2 & \gamma_1 & 0\\
  \end{array}
\right)$$ then the system takes the form
$$
\displaystyle \left\{
  \begin{array}{ccllccl}
    \dot{\omega_1} & = & -m \omega_2 \omega_3-\gamma_2,&&\dot{\gamma_1} & = & \gamma_2\omega_3-\gamma_3\omega_2,\\
    \dot{\omega_2} & = & m \omega_1 \omega_3+\gamma_1,&&\dot{\gamma_2} & = & \gamma_3\omega_1-\gamma_1\omega_3,\\
    \dot{\omega_3} & = & 0,&&\dot{\gamma_3} & = & \gamma_1\omega_2-\gamma_2\omega_1.\\
  \end{array}
\right. $$ 
It is a two degrees of freedom integrable Hamiltonian system with Poisson
structure
\begin{equation}
\label{eq:lagrange top bis}
  \begin{array}{c||cccccc}
    \{.,.\} & \omega_1 & \omega_2 & \omega_3 & \gamma_1 & \gamma_2 & \gamma_3 \\
    \hline \hline
    \omega_1 & 0 & -(1+m)\omega_3 & \omega_2/(1+m) & 0 & -\gamma_3 & \gamma_2\\
    \omega_2 & (1+m)\omega_3 & 0 & -\omega_1/(1+m) & \gamma_3 & 0 & -\gamma_1\\
    \omega_3 & -\omega_2/(1+m) & \omega_1/(1+m) & 0 & -\gamma_2/(1+m) & \gamma_1/(1+m) & 0\\
    \gamma_1 & 0 & -\gamma_3 & \gamma_2/(1+m) & 0 & 0 & 0\\
    \gamma_2 & \gamma_3 & 0 & -\gamma_1/(1+m) & 0 & 0 & 0\\
    \gamma_3 & -\gamma_2 & \gamma_1/(1+m) & 0 & 0 & 0 & 0\\
  \end{array}
\end{equation}
and Hamiltonian
$$H=\frac{1}{2}\left( \omega_1^2+\omega_2^2+(1+m)
\omega_3^2 \right)-\gamma_3.$$ 
The second first integral is $$H_{-1}=(1+m)\omega_3.$$ and the Casimir
functions are
$$H_1=-\omega_1\gamma_1-\omega_2\gamma_2-(1+m)\omega_3\gamma_3 ,$$
$$H_2=\frac{1}{2}\left( \gamma_1^2+\gamma_2^2+
\gamma_3^2\right).$$ The spectral curve $\widetilde{C}$ is given
by
\begin{equation}
\label{eq:courbe spectrale lagrange top}
y^2-f(x)=y^2-x^4-2h_{-1}x^3-\left(2h+\frac{m}{1+m}h_{-1}^2\right)x^2-2h_1x-2h_2=0.
\end{equation}

\begin{figure}
\centering
\includegraphics[width=7cm,height=5cm]{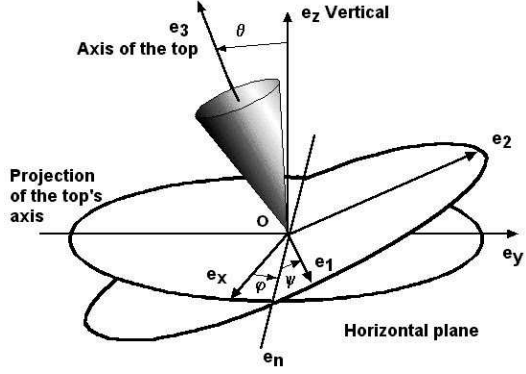}
\caption{the Lagrange top}
\label{lagrangetop}
\end{figure}

The system (\ref{eq:lagrange top bis}) describes the motion of a symmetric
rigid body spinning about its  axis whose base point is fixed ((fig.
\ref{lagrangetop}). A constant vertical gravitational force acts on the center
of mass of the top, which lies on its axis. 
The vector $\gamma $ is the unit vector $e_z$ expressed in body coordinates,
while the vector $\omega $ is the angular velocity of the body. For more
details we refer the reader to \cite{arnold,cushman2}. 
For completeness we
give below the Lagrangean function in Euler coordinates 
$\phi,\psi,\theta$ (shown of fig.
\ref{lagrangetop}), which are local coordinates on an open subset of the
configuration space $SO(3)$.
This problem will have three degrees of freedom. It has three first
integrals : the total energy $E$, the projection
$M_z$ of the angular momentum on the vertical,  the projection $M_3$ of the
angular momentum vector on the $e_3$ axis (figure
\ref{lagrangetop}).

\begin{figure}
\centering
\includegraphics[width=12cm,height=6cm]{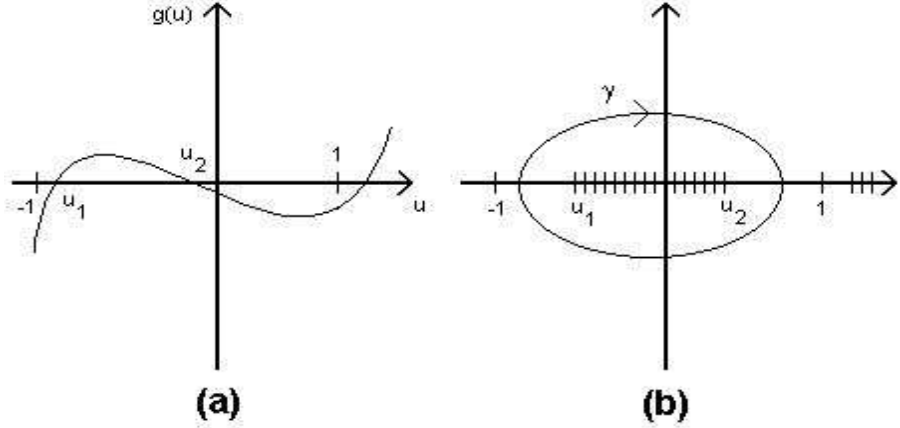} 
\caption{{\bf (a)} graph of the function $g(u)$ and
{\bf (b)} projection of the cycle $\gamma$ on the $u$-plane}
\label{graphcycle}
\end{figure}

Let $A=B\neq C$ be the moments of inertia of the body at $0$, and
let $e_1,e_2$ and $e_3$ the unit vectors of a right moving
co-ordinate system connected to the body, directed along the
principal axes at fixed point $0$. We note by $\omega$ the angular
velocity of the top which is expressed in terms of the derivates
of the Euler angles by the formula (cf \cite{arnold})
$$\omega=\omega_1 e_1+\omega_2 e_2+\omega_3
e_3=\dot{\theta}e_1+(\dot{\phi}\sin\theta
)e_2+(\dot{\psi}+\dot{\phi}\cos\theta )e_3$$ where $0<\phi<2\pi,
0<\psi<2\pi$ and $0<\theta<\pi$. Since
$T=\frac{1}{2}(A\omega_1^2+B\omega_2^2+C\omega_3^2)$, the kinetic
energy is given
$$T=\frac{A}{2}(\dot{\theta}^2+\dot{\phi}^2\sin^2\theta)+\frac{C}{2}(\dot{\psi}+\dot{\phi}\cos\theta)^2$$
and the potential energy is equal to $$U={\bf m}gl\cos\theta$$
where $l$ is the distance between the fixed point and the center
of mass of the top. The Lagrangian function reads
$$L=T-U=\frac{A}{2}(\dot{\theta}^2+\dot{\phi}^2\sin^2\theta)+\frac{C}{2}(\dot
{\psi}+\dot{\phi}\cos\theta)^2-{\bf m}gl\cos\theta.$$ Let
$p_\phi,p_\psi$ and $p_\theta$ be the conjugate moments. To the
cyclic co-ordinates $\phi$ and $\psi$ correspond the first
integrals $$p_\phi=\frac{\partial L}{\partial
\dot{\phi}}=M_z=\dot{\phi}(A\sin^2\theta+C
\cos^2\theta)+\dot{\psi}C\cos\theta,$$ $$p_\psi=\frac{\partial
L}{\partial
\dot{\psi}}=M_3=\left(\dot{\phi}\cos\theta+\dot{\psi}\right) C.$$
The last conjugate moment $p_\theta$ is equal to $p_\theta=A
\dot{\theta}.$ The momentum mapping of the Lagrange top is $$F:
T^{*}V \longrightarrow \bbbr^3$$
$$(\phi,\psi,\theta,p_\phi,p_\psi,p_\theta) \longmapsto
(E,M_3,M_z)$$ Eliminating $\dot{\phi}$ and $\dot{\psi}$, we get
the total energy $E$ of the system as
$$E=\frac{1}{2A}{p_{\theta}}^2+\frac{M_3^2}{2C}+\frac{(M_z-M_3
\cos\theta)^2}{2A\sin^2\theta}+{\bf m}gl\cos\theta.$$ Let
$$a_1=\frac{2M_3}{A},a_2=\frac{2E}{A}+\frac{M_3^2}{A}\left(\frac{1}{A}-
\frac{1}{C}\right),a_3=\frac{2M_z}{A}.$$ and obviously
$(E,M_3,M_z) \longrightarrow (a_1,a_2,a_3)$ is a bi-polynomial
map. Moreover we shall assume that $$A={{\bf m} gl}.$$
Then action variables are obviously given by \cite{aksenenkova} :
$$I_1=\frac{A}{2\pi}\oint_{\gamma}\frac{\sqrt{g(u)}}{1-u^2}du,I_2=M_3,I_3=
M_z$$
where $$g(u)=2u^3-a_2u^2+(a_1a_3/2-2)u+a_2-(a_1^2+a_3^2)/4$$
and the cycle $\gamma$ is defined on figure \ref{graphcycle}.b.
It is well known \cite{whittaker} that for a real motion of Lagrange top, 
the polynomial $g(u)$ has
exactly two real roots
$u_1$ and $u_2$ on the interval $-1 \leq u \leq 1$ and one for $u>1$ 
(figure \ref{graphcycle}.a). 
The linear change of variables
$$u=-2\xi+\frac{a_2}{6},v=-2\ii\eta.$$ transforms the curve
$\Gamma$ (*) to the curve $$\Gamma'=\{\eta^2=4\xi^3-i\xi-j\}$$
where
$$i=1-\frac{a_1a_3}{4}+\frac{a_2^2}{12},j=\frac{a_2}{6}+\frac{a_1a_2a_3}{48}-\frac{a_1^2}{16}-\frac{a_2^3}{21
6}-\frac{a_3^2}{16}.$$ Subsequently we shall consider two elliptic
curves $\Gamma'$ and $$C=\{(x,y)\in \bbbc^2 :\quad
y^2=x^4+a_1x^3+a_2x^2+a_3x+1\}.$$

\begin{remark} $C$ is nothing but the curve $\tilde{C}$ (\ref{eq:courbe spectrale lagrange
top}), where
 $$A=1, \quad E=H, \quad M_3=H_{-1}, \quad
M_z=H_2, \quad m=C-1.$$
\end{remark}

\begin{figure}
\centering
\includegraphics[width=14.4cm,height=5.1cm]{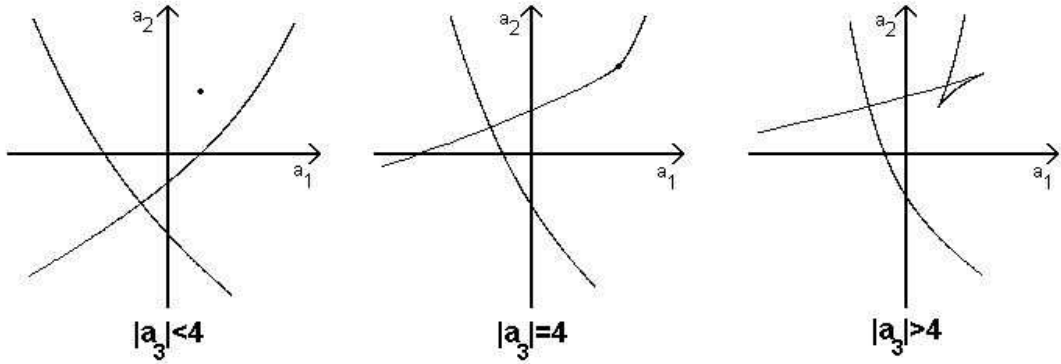}
\caption{the discriminant locus of $f(x)$} \label{disc}
\end{figure}

\begin{figure}
\centering
\includegraphics[width=5.8cm,height=5.7cm]{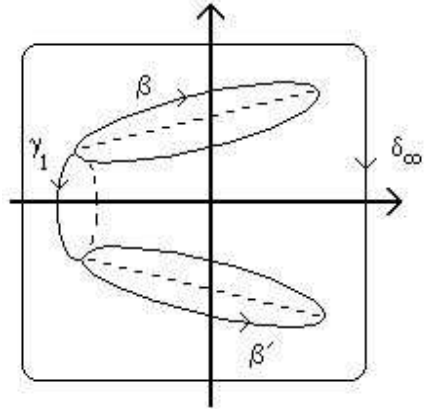}
\caption{projection of the cycles $\gamma_1,\delta_1,\delta_1'$ and $\gamma_\infty$ on the
$x$-plane} \label{cycles}
\end{figure}

The curves $C$ and $\Gamma$ are isomorphic,
more precisely
$\Gamma'$ as the Jacobian
$J(C)$ of $C$ \cite{weil2}.
The birational mapping identifying $C$ and $\Gamma'$ is given by
$$(x,y) \longmapsto \left(\xi=\frac{A_1}{x-r_0}+\frac{A_2}{2},\eta=\frac{ y A_1 }
{(x-r_0)^2 }\right) \eqno{(**)} $$ where $r_0$ is a root of $f(x)$
such that its real part is positive  and $A_1=r_0^3+\frac{3}{4}a_1
r_0^2+\frac{1}{2}a_2 r_0+\frac{1}{4}a_3,A_2=r_0^2+\frac{1}{2}a_1
r_0+\frac{1}{6}a_2$. The map (**) sends the root $r_0$ to $\infty$
and then translates the barycenter of the three remaining roots
into the origin \cite{bateman}. Using (**) it is easy to check
$$\frac{d\xi}{\eta}=-\frac{dx}{y}. \eqno{(***)} $$
Now we are going to study the discriminant locus $\Delta \subset
\bbbr^3$ of the polynomial $f(x)=x^4+a_1x^3+a_2x^2+a_3x+1$. We
denote $\Delta_{c}=\Delta \cap \{a_3$=c$\} \subset \bbbr^2$ in the
$(a_1,a_2)$-plane. Let us consider the following cases :
\begin{itemize}
\item If $f(x)$ has a real double root $u$ then
$$f(x)=(x-u)^2(x^2+\alpha x+\beta), \alpha \in \bbbr,
\beta,u \in \bbbr \backslash \{0\}.$$
Hence
$$ \left\{ \begin{array}{c}
a_1=(c+{2}/{u})/u^2-2u\\
a_2={-3}/{u^{2}}-{2c}/{u}+u^{2}
\end{array} \right.$$
$\Delta_{c}$ is parameterized by $u \in \bbbr
\backslash
\{0\}$.
\item If $f(x)$ has a real triple root $u$ then
$$f(x)=(x-u)^3(x-\alpha), \alpha,u \in \bbbr \backslash
\{0\}.$$
\begin{itemize}
\item If $c=\pm4$ then $u=\mp 1$ is a real quadruple root. It is the point $(a_1,a_2)=(\pm 4,6)$.
\item If $|c|>4$ then there are two possibilities for $u$, moreover $u$ has
the sign of
$-c$.
\item If $|c|<4$ then $f$ can not have a real triple root.
\end{itemize}
\item If $f(x)$ has two double roots then
$$f(x)=(x^2+\alpha x+\beta)^2, \alpha \in \bbbr,
\beta \in \bbbr \backslash \{0\}.$$
\begin{itemize}
\item If $(\alpha,\beta)=(-c/2,-1)$ then $f$ have two real distinct double
roots of opposite sign. Therefore the two branchs of $\Delta_{c}$
have an intersection point at $(a_1,a_2)=(-c,-2+c^2/4)$.
\item If $(\alpha,\beta)=(c/2,1)$ then
\begin{itemize}
\item If $|c|>4$ then we have two different real double roots of the
same sign as $-c$. They represent a normal crossing of $\Delta_{c}$ with
coordinates
$(a_1,a_2)=(c,2+c^2/4).$
\item If $|c|<4$ then we have a pair of complex conjugate  double roots.
They represent  an isolated point of the real discriminant locus with
coordinates
$(a_1,a_2)=(c,2+c^2/4)$.
\end{itemize}
\end{itemize}
\end{itemize}
The sections $\Delta_{c}$ of the dicriminant locus $\Delta $ are shown on
figure \ref{disc}. Let $\C_{c}$ be the connected
component of the complement to $\Delta_{c}$ in $\bbbr^2$, in which
$f(x)$ has no real root. $$\C=\left\{(a_1,a_2,a_3) \in
\bbbr^3 : \quad (a_1,a_2) \in \C_{a_3} \mbox{ and } |a_3|<4
\right\}$$
\begin{lemme}
We have $$I_1=\frac{A\ii}{2\pi}\oint_{\gamma_1}\frac{y}{x^2}dx$$
where $y^2=x^4+a_1x^3+a_2x^2+a_3x+1$ and the cycle $\gamma_1$ is
defined on figure \ref{cycles}.
\end{lemme}

{\bf Proof }
We have
\[ \frac{\partial I_1}{\partial
a_2}=\frac{A}{4\pi}\oint_{\gamma}\frac{du}{v}=-\frac{A\ii}{4\pi}\oint_{\gamma}\frac{d\xi}{\eta}
\stackrel{(***)}{=}
\frac{A\ii}{4\pi}\oint_{\gamma_1}\frac{dx}{y}=\frac{A\ii}{2\pi}\frac{\partial
}{\partial a_2} \left(\oint_{\gamma_1}\frac{y}{x^2}dx \right) .\]
Then
$$I_1=\frac{A\ii}{2\pi}\oint_{\gamma_1}\frac{y}{x^2}dx+g(a_1,a_3)$$
where $g(a_1,a_3)$ is a function. To compute $g(a_1,a_3)$, we
note that for any fixed $(a_1,a_3)$ such that the polynomial
$f(x)$ has no real root, we may continuously deform $a_2$ in such
a way, that $(a_1,a_2,a_3)$ lies on $\Delta$. But under such a
deformation the cycle $\gamma_1(a_1,a_2,a_3)$ vanishes. And hence
$I_1(a_1,a_2,a_3)=0$ and $\oint_{\gamma_1}{y}dx/{x^2}=0$ which
implies $g(a_1,a_3)=0$.

\subsubsection{The monodromy of Lagrange top}
\begin{figure}
\centering
\includegraphics[width=5.7cm,height=6.1cm]{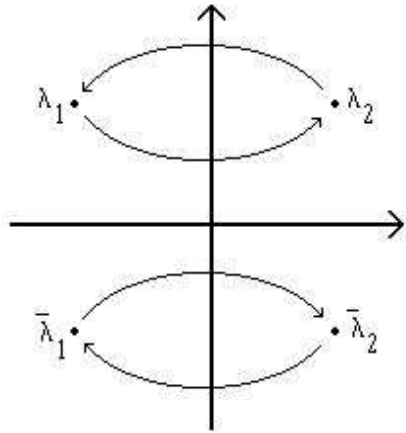}
\caption{the $x$-plane}
\label{lambdaplane}
\end{figure}

Let $F: T^*V \longrightarrow \bbbr^3$ be the moment map of the Lagrange top,
where $V=SO(3)$. We consider the
fibration
$$\widetilde{F} : T^*V \backslash F^{-1}(\Delta ) \longrightarrow
\bbbr^3 \backslash \Delta .$$ This is a proper topological fibration,
the fibers of which are diffeomorphic to three-tori ${\T}^3$. We consider
the real monodromy of $\widetilde{F}$ defined as the action of
$\pi_1(\bbbr^3 \backslash \Delta ,c)$ on
$H_1(\widetilde{F}^{-1}(c),\bbbz)$, $c=(c_1,c_2,c_3) \in \bbbr^3
\backslash \Delta $. We choose now a basis $\alpha_1,\alpha_2$ and
$\alpha_3$ of $H_1(\widetilde{F}^{-1}(c),\bbbz)$ in the following
way :
\begin{itemize}
\item For $\alpha_1$ we take the path on $\widetilde{F}^{-1}(c)$ defined by fixing $\phi,\psi$. 
$\theta,p_\theta$ make one circle on the curve defined by the equation
$$c_1=\frac{1}{2A}{p_{\theta}}^2+\frac{c_2^2}{2C}+\frac{(c_3-c_2
\cos\theta)^2}{2A\sin^2\theta}+{\bf m}gl\cos\theta$$
\item For $\alpha_2$ we fix $\theta,p_\theta$ and $\phi$ and $\psi$ run through
the interval $[0,2\pi]$.
\item For $\alpha_3$ we fix $\theta,p_\theta$ and $\psi$ and $\phi$ run through the
interval $[0,2\pi]$.
\end{itemize}
With such a choice of basis of $H_1(\widetilde{F}^{-1}(c),\bbbz)$,
the action variables are given by
$$I_i=\frac{1}{2\pi}\oint_{\alpha_i} \sigma, i=1,2,3.$$ where
$\sigma=p_\theta d\theta +p_\phi d\phi+p_\psi d\psi$ is the
fundamental one-form on $T^*V$.\\
\\
{\bf Theorem (R. Cushman) \cite{duistermaat,cushman2} } {\it If $z_0 \in \C$
then
$\pi_1(\bbbr^3
\backslash D,z_0)=\bbbz$ and the real monodromy of $F$ can be represented, on
the basis $\alpha_i$ (defined above) for $H_1(F^{-1}(c),\bbbz)$,
by the matrix
$$\left( \begin{array}{ccc}
1 & 0 & 0\\
1 & 1 & 0\\
0 & 0 & 1\\
\end{array} \right).$$ }
\\
{\bf Proof }
The proof of this theorem will follow from the following elementary
\begin{lemme}
The real discriminant locus of the real polynomial
$f(x)=(x^2+1)^2+(a_1x+a_2)x^2$ in a small neighborhood of the origin in
$\bbbr^2\{a_1,a_2\}$ consists of the point $(0,0)$.
When $(a_1,a_2)$ makes one turn around $(0,0)$ in a negative direction then the roots of
$f(x)$ exchange their places as it is shown on figure \ref{lambdaplane}.
\end{lemme}

\paragraph{Proof }The proof is straightforward.
\begin{remark}
For $(a_1,a_2)\in \bbbr^2$ sufficiently small, the real polynomial
$f(x)$ has either two double roots or it has no double root at
all. Hence the real discriminant locus of $f(x)$ is of codimension
two and hence it is the point $(0,0)$. This phenomenon has a more general
nature, see Looijenga \cite{looijenga}.
\end{remark}

To compute the monodromy of the action variables (equivalently, the monodromy
of the homology bundle of the Lagragian fibration $\tilde{F}$), we shall
consider  the monodromy of the homology  bundle of the Milnor fibration 
$\B$ of the polynomial $y^2-x^4-a_1x^3-a_2 x^2-a_3x-1$. This
is a fibration 
with fiber
$\widetilde{C}$ over $\bbbr^3 \backslash \Delta$, defined by 

$$\B \longrightarrow \bbbr^3 \backslash \Delta$$ 
$$\{y^2=x^4+a_1x^3+a_2 x^2+a_3x+1\} \longmapsto (a_1,a_2,a_3).$$
$\pi_1(\bbbr^3\backslash \Delta,z_0)$ is not trivial if and only if 
$z_0 \in \C$.

\begin{figure}
\centering
\includegraphics[width=5.6cm,height=6.1cm]{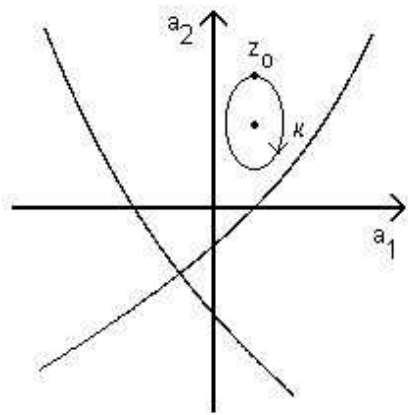}
\caption{the loop $\kappa \in \pi_1(\C_c,z_0)$}
\label{loopk}
\end{figure}

Denote $P_0=(c,2+c^2/4)$ on the $(a_1,a_2)$-plane, and
consider a simple negatively oriented (because the map
$(E,M_3,M_z) \longrightarrow(a_1(E,M_3,M_z),a_2(E,M_3,M_z),a_3(E,M_3,M_z))$ reverse the
orientation) loop $\kappa$ around $P_0$, figure \ref{loopk}.

\begin{figure}
\centering
\includegraphics[width=12.9cm,height=5.3cm]{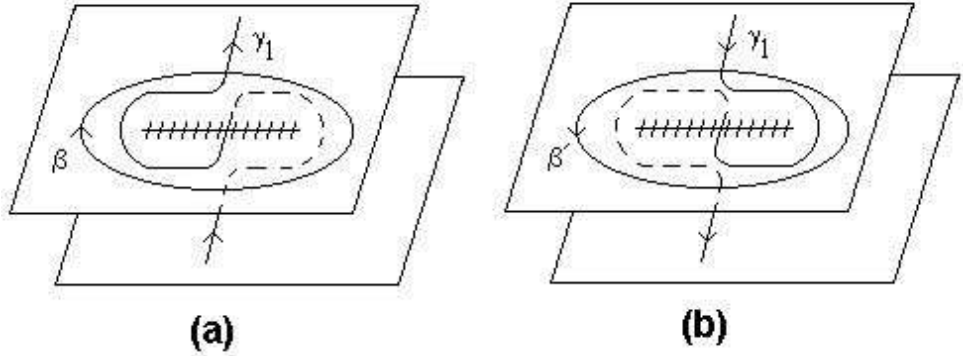}
\caption{the cycle $\gamma_1$ {\bf (a)} and {\bf (b)}}
\label{cyclegam1}
\end{figure}

This defines $\kappa$ as a loop in $\bbbr^3 \backslash \Delta$
with $z_0 \in \C$ as base point. It is possible to deform
continuously $\kappa$ to a loop (with the same orientation)
contained in $\C \cap \{a_3=0\}$. The monodromy of roots of $f(x)$
induces the monodromy of cycles in $H_1(\widetilde{C},\bbbz)$.
This situation is described in figures \ref{cyclegam1}.a and
\ref{cyclegam1}.b.
Let $\gamma_1'$ be the image of $\gamma_1$ after making one turn
along $\kappa$ in negative direction. Then 
the classical Picard-Lefschetz formula \cite{agv} implies
$\gamma_1' \equiv
\gamma_1+\delta_1 -\delta_1'$ and moreover we have
$\gamma_{\infty} \equiv \delta_1 -\delta_1'$ where the projections
of $\delta_1,\delta_1'$ and $\gamma_{\infty}$ on the $x$-plane are
shown on figure \ref{cycles}. That is to say
$$I_1'=\frac{A\ii}{2\pi}\oint_{\gamma_1'}\frac{\sqrt{f(x)}}{x^2}dx=\frac{A\ii}
{2\pi}\oint_{\gamma_1}\frac{\sqrt{f(x)}}{x^2}dx+\frac{A\ii}{2\pi}
\oint_{\gamma_\infty}\frac{\sqrt{f(x)}}{x^2}dx$$ and
$$\oint_{\gamma_\infty}\frac{\sqrt{f(x)}}{x^2}dx=2\ii\pi \mbox{
residue}_{x=\infty}\left(\frac{\sqrt{f(x)}}{x^2}dx \right)=-\ii\pi
a_1.$$ We see that $I_1$ is transformed to $I_1+I_2$.\\

\subsection{The case $g=2$}

For this case the system (\ref{eq:system}) is
\begin{equation}
\label{eq:case g=2}
\frac{d}{dt}\left( \chi \lambda +\Gamma_0-\Gamma_1\lambda^{-1}-\Gamma_2\lambda^{-2} \right)=
\left[\chi \lambda
+\Gamma_0-\Gamma_1\lambda^{-1}-\Gamma_2\lambda^{-2},\chi \lambda +\Omega \right].
\end{equation}
As above we put
$$\Gamma_1=
\left(
  \begin{array}{ccc}
    0 & -\gamma_3 & \gamma_2\\
    \gamma_3 & 0 & -\gamma_1\\
    -\gamma_2 & \gamma_1 & 0\\
  \end{array}
\right), \quad
\Gamma_2=\left(
  \begin{array}{ccc}
    0 & -\theta_3 & \theta_2\\
    \theta_3 & 0 & -\theta_1\\
    -\theta_2 & \theta_1 & 0\\
  \end{array}
\right). $$ 
In these notations
$$\frac{d}{dt}\Gamma_0=\left[\Gamma_0,\Omega
\right]-\left[\Gamma_1,\chi \right],$$
$$\frac{d}{dt}\Gamma_1=\left[\Gamma_1,\Omega
\right]+\left[\Gamma_2,\chi \right],$$
$$\frac{d}{dt}\Gamma_2=\left[\Gamma_2,\Omega \right].$$ or also
 $$
\displaystyle \left\{
  \begin{array}{ccllccllccl}
    \dot{\omega_1} & = & -m \omega_2 \omega_3-\gamma_2,& &\dot{\gamma_1} & = & \gamma_2\omega_3-\gamma_3\omega_2+\theta_2, & &
     \dot{\theta_1} & = & \omega_3\theta_2-\omega_2\theta_3,\\
    \dot{\omega_2} & = & m \omega_1 \omega_3+\gamma_1, & & \dot{\gamma_2} & = & \gamma_3\omega_1-\gamma_1\omega_3-\theta_1, & &
    \dot{\theta_2} & = & \omega_1\theta_3-\omega_3\theta_1,\\
    \dot{\omega_3} & = & 0, & &\dot{\gamma_3} & = & \gamma_1\omega_2-\gamma_2\omega_1, & &\dot{\theta_3} & = & \omega_2\theta_1-\omega_1\theta_2.\\
  \end{array}
\right. $$ Let us consider this Poisson structure $$ \displaystyle
\begin{array}{c||ccccccccc}
    \{.,.\} & \omega_1 & \omega_2 & \omega_3 & \gamma_1 & \gamma_2 & \gamma_3 & \theta_1 &
    \theta_2 & \theta_3\\

    \hline \hline

    \omega_1 & 0 & -(1+m)\omega_3 & \frac{\omega_2}{1+m} & 0 & -\gamma_3 & \gamma_2 & 0 & -\theta_3 & \theta_2\\

    \omega_2 & (1+m)\omega_3 & 0 & -\frac{\omega_1}{1+m} & \gamma_3 & 0 & -\gamma_1 & \theta_3 & 0 & -\theta_1\\

    \omega_3 & -\frac{\omega_2}{1+m} & \frac{\omega_1}{1+m} & 0 & -\frac{\gamma_2}{1+m} & \frac{\gamma_1}{1+m}
    & 0 & -\frac{\theta_2}{1+m} & \frac{\theta_1}{1+m} & 0\\

    \gamma_1 & 0 & -\gamma_3 & \frac{\gamma_2}{1+m} &  0 & \theta_3 & -\theta_2 & 0 & 0 & 0\\

    \gamma_2 & \gamma_3 & 0 & -\frac{\gamma_1}{1+m} &  -\theta_3 & 0 & \theta_1 & 0 & 0 & 0\\

    \gamma_3 & -\gamma_2 & \frac{\gamma_1}{1+m} & 0 & \theta_2 & -\theta_1 & 0 & 0 & 0 & 0\\

    \theta_1 & 0 & -\theta_3 & \frac{\theta_2}{1+m} & 0 & 0 & 0 &
    0 & 0 & 0\\

    \theta_2 & \theta_3 & 0 & -\frac{\theta_1}{1+m} & 0 & 0 & 0 &
    0 & 0 & 0\\

    \theta_3 & -\theta_2 & \theta_1 & 0 & 0 & 0 & 0 &
    0 & 0 & 0\\

  \end{array}
$$
The Hamiltonian function corresponding to (\ref{eq:case g=2}) is
$$H=\frac{1}{2}\left( \omega_1^2+\omega_2^2+(1+m)
\omega_3^2 \right)-\gamma_3.$$
The Hamiltonian functions in involution with $H$ are
$$H_{-1}=(1+m)\omega_3,$$
$$H_1=-\omega_1\gamma_1-\omega_2\gamma_2-(1+m)\omega_3\gamma_3-\theta_3 .$$
The Casimir functions are
$$H_2=\frac{1}{2}\left( \gamma_1^2+\gamma_2^2+ \gamma_3^2\right)
 -\omega_1\theta_1-\omega_2\theta_2 -(1+m)\omega_3\theta_3
,$$
$$H_3=\gamma_1\theta_1+\gamma_2\theta_2+\gamma_3\theta_3 ,$$
$$H_4= \frac{1}{2}\left(\theta_1^2+\theta_2^2+\theta_3^2\right).$$

The spectral curve $\widetilde{C}$ is given by
$$y^2=x^6+2h_{-1}x^5+\left(2h+\frac{m}{1+m}h_{-1}^2\right)x^4+2h_1x^3+2h_2x^2+2h_3x+2h_4.$$
The monodromy of cycles on spectral curve $\widetilde{C}$
generates the monodromy of momentum map associated to the system
(\ref{eq:case g=2}).

\subsubsection{The discriminant of $(x^2+1)^3+x^3(ax^2+bx+c)$}
Let us consider the real discriminant $\Delta(a,b,c)$ of the
polynomial $P(x)=(x^2+1)^3+x^3(ax^2+bx+c)$ when $(a,b,c)$ is
closed to $(0,0,0)$. Assume that
$$P(x)=\left ({x}^{2}+{ c_1}\,x+{ c_2}\right )^{2}\left ({x}^{2}+{
d_1} \,x+{ d_ 2}\right )$$
and hence
$$ \left\{
\displaystyle
  \begin{array}{ccl}
    a & = & 2\alpha(c_2-1)(c_2^3-1)/c_2^3\\
    b & = & (c_2-1)^3(c_2^3+3c_2^2+3c_2+5)/3c_2^2\\
    c & = & 2\alpha(c_2-1)(2c_2^3+3c_2-5)/3c_2^2\\
    c_1 & = & \alpha (c_2-1)\\
    d_1 & = & -2\alpha(c_2-1)/c_2^3\\
    d_2 & = & c_2^{-2}\\
  \end{array} \right.$$
where $\alpha$ verifies $3\alpha^2=c_2(c_2+2)$ and $c_2\neq 0$.
The discriminant of $\left ({x}^{2}+{ c_1}\,x+{ c_2}\right )^{2}$
is
$$\Delta_1(c_2)=c_1^2-4c_2= 1/3\,{ c_2}\,\left (-10+{{ c_2}}^{3}-3\,{ c_2}\right )
$$
and the discriminant of $\left ({x}^{2}+{ d_1}
\,x+{ d_2}\right )$ is
$$\Delta_2(c_2)=d_1^2-4d_2=-4/3\,{\frac {2\,{{ c_2}}^{3}+3\,{ c_2}-2}{{{ c_2}}^{5}}}
$$

\begin{figure}
\centering
\includegraphics[width=9.3cm,height=6.8cm]{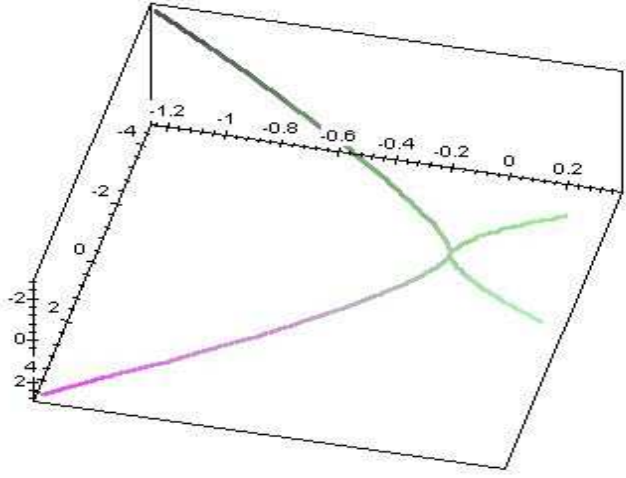}
\caption{discriminant of $f(x)$}
\label{fig:dicr}
\end{figure}

It is easy to check that $\Delta_1(c_2)$ and $\Delta_2(c_2)$ are
negative when $c_2$ is close to $1$.Therefore the
discriminant $\Delta(a,b,c)$ is parameterized near $(0,0,0)$ by
$$ \left\{
\displaystyle
  \begin{array}{ccl}
    a & = & 2\alpha(c_2-1)(c_2^3-1)/c_2^3\\
    b & = & (c_2-1)^3(c_2^3+3c_2^2+3c_2+5)/3c_2^2\\
    c & = & 2\alpha(c_2-1)(2c_2^3+3c_2-5)/3c_2^2, c_2\in (0,\infty)\\
    \end{array} \right.$$
(see  figure \ref{fig:dicr}).
Denote the set on figure \ref{fig:dicr} by $\tilde{\Delta }$. The above shows that the
connected component of the
complement to the discriminant locus, in which  the polynomial $(x^2+1)^3+x^3(ax^2+bx+c)$ has no
real roots is homeomorphic to $\bbbr^3 \backslash \tilde{\Delta } $. Moreover this implies that,
more generally, the connected component $\C\subset \bbbr^6$ of the complement to the discriminant
locus in which the spectral polynomial 
$$x^6+2h_{-1}x^5+\left(2h+\frac{m}{1+m}h_{-1}^2\right)x^4+2h_1x^3+2h_2x^2+2h_3x+2h_4$$
has no real roots, is homeomorphic to $(\bbbr^3 \backslash \tilde{\Delta })\times \bbbr^3$.
Therefore we have the following
\begin{lemme}
The fundamental group of $\C$ is a free group with three generators.
\end{lemme}

\subsubsection{The monodromy of the generalized Lagrange top}

The monodromy group of the top is a homomorphism from $\pi _1(\C,p_1)$ to $Aut(\bbbz^3)$, 
where $\bbbz^3=H_1({\T}^3,\bbbz)$.

\begin{figure}
\centering
\includegraphics[width=7.5cm,height=5.7cm]{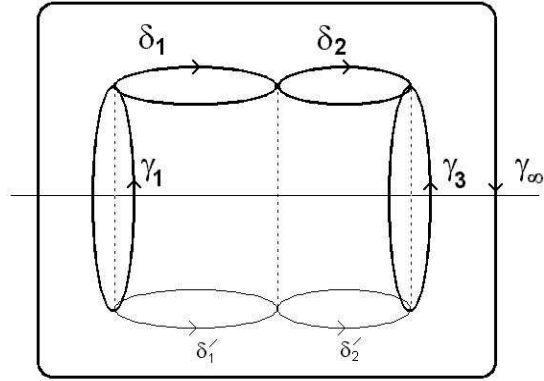}
\caption{projection of the cycles $\gamma_1,\gamma_3,\gamma_\infty,\delta_1,\delta_2$ on the $x$-plane$$}
\label{fig:cycles g=2}
\vspace{0.5cm}
\end{figure}

Consider the basis
$\{\gamma_1,\gamma_3,\gamma_\infty,\delta_1,\delta_2\}$ 
 of $H_1(\widetilde{C},\bbbz)$ shown on
(fig
\ref{fig:cycles g=2}). The
cycles generating the  Liouville tori are the cycles
$\gamma_1,\gamma_3,\gamma_\infty$.

\begin{figure}
\centering
\includegraphics[width=6.5cm,height=6.5cm]{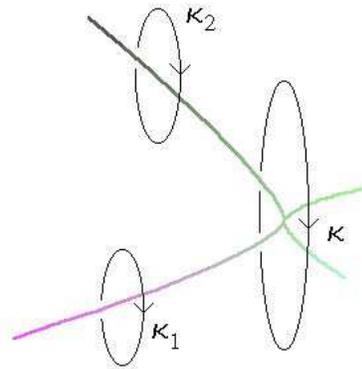}
\caption{the loops $\kappa$}
\label{fig:kappa}
\end{figure}

\begin{figure}
\centering
\includegraphics[width=6.5cm,height=4.1cm]{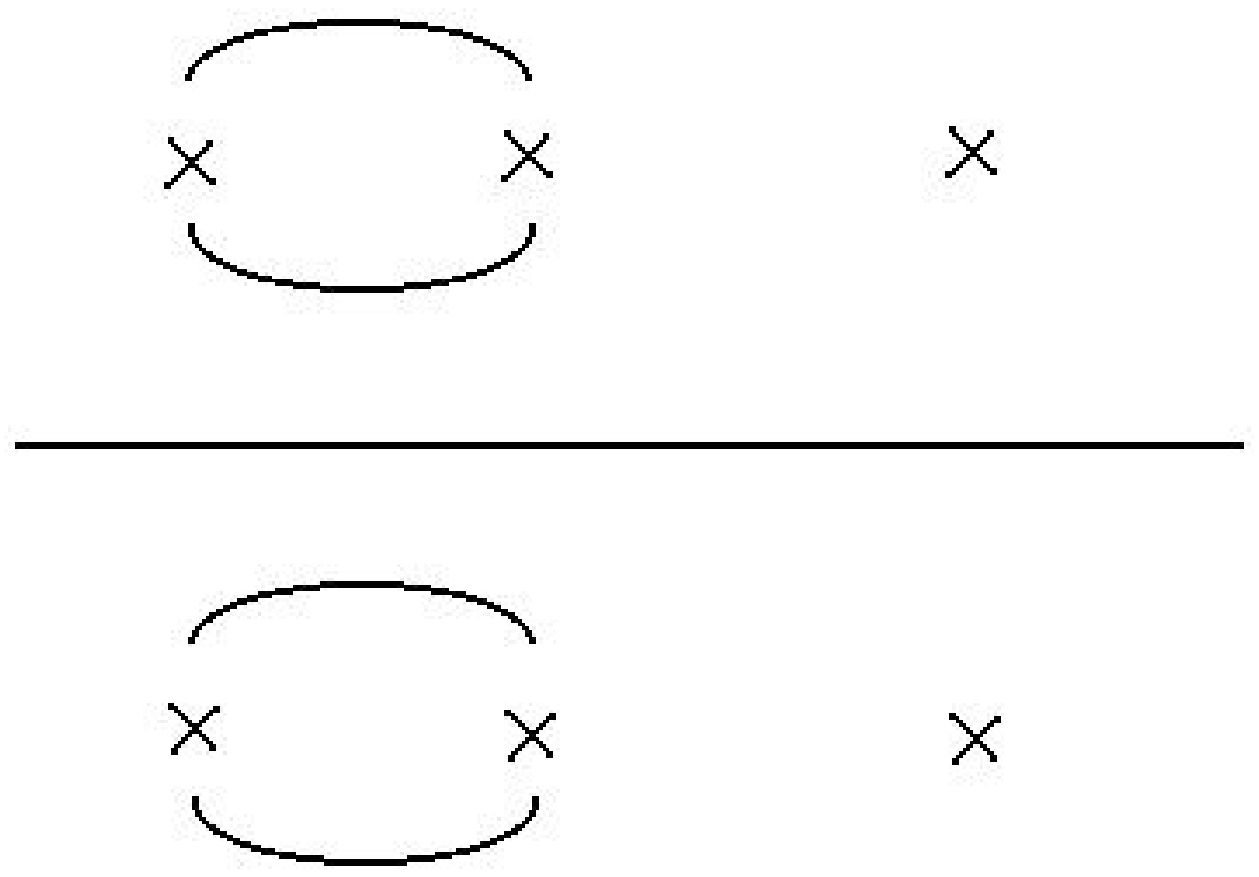}
\caption{monodromy of the roots of $f(x)$} \label{fig:mvtracines1}
\end{figure}

Let $\kappa_1 \in \pi _1(\C, p_1)$ be the loop 
shown on figure \ref{fig:kappa}. The monodromy of the roots of the polynomial
$f(x)$, induced by this loop are shown on fig.
\ref{fig:mvtracines1}. Therefore, when $(a_1,a_2,a_3)$ makes one turn along
$\kappa$, the cycle $\gamma_1$ is transformed to
$\gamma'_1$, where
$$\gamma'_1=\gamma_1+\delta_1-\delta'_1
=\gamma_1-\gamma_3+\gamma_\infty.$$ 
The monodromy of cycles is
given by the following matrix (in the basis
$\{\gamma_1,\gamma_3,\gamma_\infty\}$) $$M_{\kappa_1}=\left(
  \begin{array}{ccc}
    1 & 0 & 0 \\
   -1 & 1 & 0\\
   1 & 0 & 1\\
  \end{array}
\right).$$

\begin{figure}
\centering
\includegraphics[width=6.5cm,height=4.1cm]{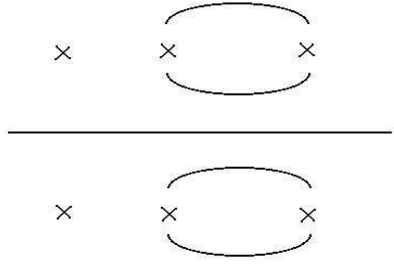}
\caption{the monodromy of the roots of $f(x)$} \label{fig:mvtracines2}
\end{figure}

Consider the loop $\kappa_2\in \pi _1(\C, p_2)$  shown on figure \ref{fig:kappa}.
The monodromy of the roots of the polynomial
$f(x)$ induced by  $\kappa_2$ is shown on figure
\ref{fig:mvtracines2}. The cycle $\gamma_1$  is transformed to
$\gamma'_1$ where
$$\gamma'_3=\gamma_3+\delta_2-\delta'_2
=\gamma_3-\gamma_1+\gamma_\infty.$$ 
The monodromy of the cycles is
given by the following matrix (in the basis
$\{\gamma_1,\gamma_2,\gamma_\infty\}$) $$M_{\kappa_2}=\left(
  \begin{array}{ccc}
    1 & -1 & 0 \\
    0 & 1 & 0\\
    0 & 1 & 1\\
  \end{array}
\right).$$

In a similar way we may choose a third generator $\kappa_3$ and compute its image in $Aut(\bbbz^3)$.

%
%
%

\bibliographystyle{amsplain}

\end{document}